\documentclass[twocolumn]{aastex61}

\usepackage{subfigure}
\usepackage{url}
\usepackage{hyperref}
\usepackage{epsfig}
\usepackage{graphicx}
\usepackage{sidecap}
\usepackage{hanging}
\usepackage{color}
\usepackage{multirow}
\usepackage{enumerate}
\usepackage{natbib}
\usepackage{amssymb}
\usepackage{amsmath}

\newcommand{\sci}{Sci}
\newcommand\jatis{JATIS}

\newcommand{\kepler}{{\it Kepler}}

\providecommand{\e}[1]{\ensuremath{\times 10^{#1}}}

\newcommand{\retro}{\texttt{RETrO}}
\newcommand{\h}{H$_2$/He}
\newcommand{\water}{H$_2$O}
\newcommand{\n}{N$_2$}
\newcommand{\carbon}{CO$_2$}

\submitjournal{Accepted for publication in \textit{ApJ} on September 19, 2017}

\shorttitle{Stellar Mirages in Exoplanetary Atmospheres}
\shortauthors{Dalba}

\begin{document}

\title{Out-of-Transit Refracted Light in the Atmospheres of \\Transiting and Non-Transiting Exoplanets}

\correspondingauthor{P. A. Dalba}
\email{pdalba@bu.edu}

\author[0000-0002-4297-5506]{Paul A. Dalba}
\affil{Department of Astronomy, Boston University \\
725 Commonwealth Ave., Rm 514 \\
Boston, MA 02215-1401, USA}

\begin{abstract}
Before an exoplanet transit, atmospheric refraction bends light into the line of sight of an observer. The refracted light forms a stellar mirage---a distorted secondary image of the host star. I model this phenomenon and the resultant out-of-transit flux increase across a comprehensive exoplanetary parameter space. At visible wavelengths, Rayleigh scattering limits the detectability of stellar mirages in most exoplanetary systems with semi-major axes $\lesssim$6 AU. A notable exception is almost any planet orbiting a late M or ultra-cool dwarf star at $\gtrsim$0.5 AU, where the maximum relative flux increase is $>$50 parts-per-million. Based partly on previous work, I propose that the importance of refraction in an exoplanet system is governed by two angles: the orbital distance divided by the stellar radius and the total deflection achieved by a ray in the optically thin portion of the atmosphere. Atmospheric lensing events caused by non-transiting exoplanets, which allow for exoplanet detection and atmospheric characterization, are also investigated. I derive the basic formalism to determine the total signal-to-noise ratio of an atmospheric lensing event, with application to \kepler\ data. It is unlikely that out-of-transit refracted light signals are clearly present in \kepler\ data due to Rayleigh scattering and the bias toward short-period exoplanets. However, observations at long wavelengths (e.g., the near-infrared) are significantly more likely to detect stellar mirages. Lastly, I discuss the potential for the {\it Transiting Exoplanet Survey Satellite} to detect refracted light and consider novel science cases enabled by refracted light spectra from the {\it James Webb Space Telescope.} 
\end{abstract}


\section{Introduction} \label{sec:intro}

Electromagnetic radiation propagating through a planetary atmosphere experiences refraction. This process affects the path taken by the radiation and may alter its original trajectory. The extent to which atmospheric refraction influences incoming radiation depends on the nature of the radiation and the composition and characteristics of the planetary atmosphere. This has proven beneficial for remote sensing investigations of the atmospheres of solar system bodies during occultations of background stars, the Sun, and even spacecraft. Since the basic theory of refraction during planetary occultations was developed \citep[e.g.,][]{Radau1882,Pannekoek1903,Fabry1929}, substantial theoretical and observational efforts have ushered this field into maturity \citep[see, for instance,][and references therein]{Baum1953,Wasserman1973,Elliot1979,Hubbard1993,Elliot1996,Withers2014}.

Transits of extrasolar planets present opportunities to probe the atmospheres of bodies outside of the solar system \citep[e.g.,][]{Charbonneau2000,Seager2000,Charbonneau2002}. As in any other planetary occultation, a ray of light traversing an exoplanetary atmosphere will be influenced by atmospheric refraction. Transiting exoplanets, therefore, represent a new means of probing atmospheric processes with refraction. Not long after the discovery of HD 209458 b, \citet{Seager2000}, \citet{Hubbard2001}, and \citet{Hui2002} correctly asserted that refraction effects are unimportant for exoplanets similar to HD 209458 b. Since then, the sample of transiting exoplanets that has been studied in depth has been heavily biased toward short-period exoplanets. The result of this short-period bias is a delay in understanding how refracted light can be fully exploited to characterize exoplanetary atmospheres \citep{Deming2017}. 

\citet{Hui2002} investigated refraction and the existence of caustics in exoplanetary atmospheres with a model formalism similar to gravitational lensing. They broadly separated exoplanet parameter space into refractive and non-refractive groups but lacked a sizable sample of transiting exoplanets to place their results in context. Most of the recent investigations of atmospheric refraction in an exoplanetary context focused on its influence on transmission spectroscopy. Refraction produces ``surfaces''  in exoplanet atmospheres, thereby setting upper limits on the atmospheric pressure levels that can be sensed at mid-transit \citep{Betremieux2014,Betremieux2015,Dalba2015,Betremieux2016,Betremieux2017}. This effect would be particularly significant in observations of Earth-analog exoplanets \citep[e.g.,][]{GarciaMunoz2012a,Misra2014a}. 

In contrast to its limiting effect on transmission spectra, atmospheric refraction produces another phenomenon that has the potential to serve as a new means of atmospheric characterization. Before or after a transit, a planetary atmosphere will deflect light into the line of sight of a distant observer. The light forms a distorted secondary image of the host star, which I hereafter refer to as a \emph{stellar mirage}. In a transit light curve, this unresolved mirage creates a small increase in flux peaked at the moment before transit \citep{Sidis2010}. Although the effect is so far undetected in exoplanet transit light curves, it is responsible for Lomonosov's discovery of the Venusian atmosphere during the transit of Venus in 1761 \citep{Cruikshank1983} and is the subject of other recent studies of Venus \citep[e.g.,][]{GarciaMunoz2012b,Pere2016}. 

Only \citet{Sidis2010} and \citet{Misra2014b} have investigated the out-of-transit refracted light signal in quantitative detail. \citet{Sidis2010} demonstrated that the stellar mirage would have a crescent shape (their Fig. 1). They also derived analytic expressions for the magnitude of the flux increase as a function of projected orbital separation between the star and exoplanet. Although a useful introductory work, \citet{Sidis2010} only considered a few examples of realistic extrasolar planetary systems and used toy models to calculate the size---and therefore magnitude---of the refracted image. These choices made the problem analytically tractable, but limited the applicability of their results. \citet{Misra2014b} modeled the refracted light signal outside of transit for the purpose of prioritizing future spaced-based observations of exoplanetary atmospheres. \citet{Misra2014b} improved upon \citet{Sidis2010} by utilizing a ray tracing scheme to simulate atmospheric refraction. However, \citet{Misra2014b} also used a toy model to calculate the shape and size of the refracted image.  

What is lacking in the literature is a more accurate treatment of refracted light outside of exoplanet transits, and a comprehensive exploration of parameter space to identify the systems that are most amenable to atmospheric characterization via refracted light. I intend to provide both of these here. Additionally, I aim to understand whether or not an out-of-transit refracted light signal is likely to be present in the high-precision \kepler\ data set, either for transiting or non-transiting systems or exoplanets. 

In \S\ref{sec:retro}, I describe the ray tracing model titled Refraction in Exoplanet Transit Observations, or \retro.\footnote{\retro\ will be made publicly available at \url{https://github.com/pdalba/retro}.} In \S\ref{sec:param_space} and \S\ref{sec:results}, I conduct a comprehensive parameter space search of the out-of-transit refracted light signal at visible wavelengths. In \S\ref{sec:lens}, the potential for refraction to create atmospheric lensing events that identify non-transiting exoplanets and reveal the properties of their atmospheres is investigated. I provide the formalism to estimate the detectability of atmospheric lensing events for non-transiting exoplanets at visible wavelengths, with application to the \kepler\ data set. In \S\ref{sec:disc}, the atmospheric lensing predictions of \citet{Hui2002} are revisited. I also discuss future science applications of atmospheric refraction in exoplanet systems. Lastly, the results of this work are summarized in \S\ref{sec:conclusions}.

\section{\retro: An Out-of-Transit Refracted Light Model}\label{sec:retro}

\begin{figure}
\centering
\includegraphics[width=0.9\columnwidth]{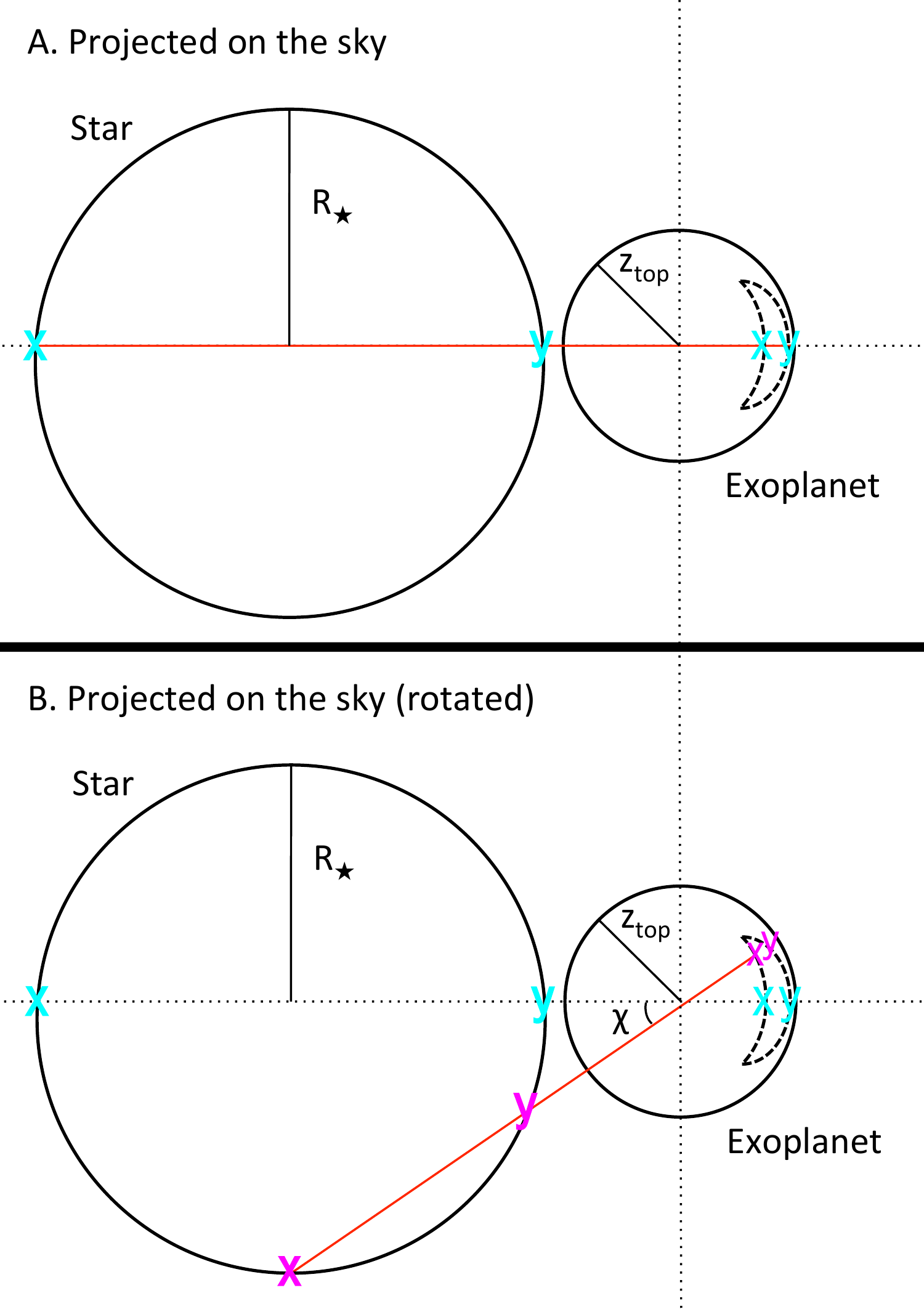}
\caption{Diagram showing the host star, exoplanet, and stellar mirage (dashed crescent) projected on the sky before or after a transit. The star, exoplanet, and mirage are not to scale. {\bf A:} Rays are traced in the equatorial plane of the star-planet system, which is designated by the red line. The blue ``x'' and ``y'' on the star map to the secondary image in the exoplanet atmosphere as shown. {\bf B:} To trace rays in three dimensions, the ray tracing plane (red line) is rotated about the center of the exoplanet through the angle $\chi$. The pink ``x'' and ``y'' on the star map to the secondary image in the exoplanet atmosphere as shown. The end result is an inverted, distorted secondary image of the stellar disk.}
\label{fig:crescent_geom}
\end{figure}

Before or after a transit, some of the light from the host star that is originally not propagating toward the observer is refracted by the exoplanetary atmosphere into the observer's line of sight. This occurs for certain rays on the entire disk of the star facing the exoplanet. The result is a coherent secondary image of the star in the exoplanet atmosphere, a \emph{stellar mirage}.\footnote{I use the terms ``secondary image'' and ``stellar mirage'' interchangeably.} The stellar mirage is an image of the entire disk of the host star, but it is distorted into a crescent shape \citep[Fig. \ref{fig:crescent_geom}, also Fig. 1 of][]{Sidis2010}. Specific intensity is conserved, so the relative flux increase measured by the distant observer is proportional to the area of the secondary image. The secondary image will always be unresolved for an exoplanet as viewed from Earth, so the effect manifests as ``shoulders'' on the transit light curve, with increases in flux prior to and after transit. 

The latitudinal extent of the crescent is primarily determined by the stellar and exoplanetary radii and the projected star-planet separation. The maximum width of the crescent is typically on the order of the atmospheric scale height because one e-folding factor in density---and also refractivity---usually provides enough bending to focus the light originating from the near and far sides of the host star.\footnote{I use the terms ``near'' and ``far'' in reference to limbs of the exoplanet and its host star. ``Near'' means the limb of the exoplanet (star) that is closest to the star (exoplanet) from the point of view of the observer. ``Far'' means the opposite.} As the projected separation between the exoplanet and star decreases, the stellar mirage becomes larger and appears at higher altitudes (lower pressures) in the exoplanetary atmosphere. The maximum effect occurs when the projected disks of the exoplanet (including its atmosphere) and host star are mutually tangent \citep{Sidis2010}. For a transiting orbit, this is the moment before any of the host star's flux reaching the observer begins experiencing attenuation from absorption, scattering, or refractive defocusing.

\subsection{Ray Tracing}

I implement the two-dimensional ray tracing scheme of \citet{Kivalov2007}. This scheme approximates the path elements of a ray as circle segments with curvature determined by the atmospheric refractivity profile. Assuming a curved path for each path element returns a second-order approximation to the actual path of light in a planetary atmosphere. Table \ref{tab:symbols} contains all of the symbols used in \retro. 

\startlongtable
\begin{deluxetable}{ll}
\tablecaption{List of symbols.\label{tab:symbols}}
\tablecolumns{2}
\tablewidth{\textwidth}
\tablehead{
\colhead{Symbol} &
\colhead{Meaning}
}
\startdata
$a$ & Exoplanet semi-major axis \\
$\mathcal{A}$ & Cauchy's coefficient of refraction \\
$A_{\rm B}$ & Bond albedo \\
$A_{\mathcal{M}}$ & Area of the stellar mirage\\
$b$ & Transit impact parameter \\
$b_{\tau=1}$ & Impact parameter of a ray with $\tau=1$\\
$\mathcal{B}$ & Cauchy's coefficient of dispersion \\
$B_{\rm HS02}$ & $B$-parameter from \citet{Hui2002} \\
$f_{\mathcal{M}}$ & Relative flux increase due to stellar mirage\\
$f_{\rm \mathcal{M},max}$ & Maximum relative flux increase \\
$g$ & Exoplanet gravitational acceleration\\
$H$ & Atmospheric pressure scale height\\
$J$ & Photoelectrons per unit time \\
$\mathcal{M}$ & Closed curve of the stellar mirage\\ 
$m_{\rm H}$ & Mass of hydrogen atom \\
$M_{\rm p}$ & Exoplanet mass\\
$n$ & Index of refraction\\
$N$ & Atmospheric particle number density \\
$P_{\rm detect}$ & Refracted light detection probability \\
$P_{\rm transit}$ & Geometric transit probability \\
$R_{\star}$ & Stellar radius\\
$s$ & Distance along ray path\\
SNR$_{\rm total}$ & Integrated SNR of lensing event\\
$t$ & Time coordinate\\
$T$ & Duration of lensing event\\
$T_{\rm atm}$ & Atmospheric temperature \\
$T_{\rm eff}$ & Stellar effective temperature\\
$T_{\rm eq}$ & Planetary equilibrium temperature\\
$X$ & Projected star-planet separation \\
$Y$ & Solar helium mass fraction \\
$z$ & Radial distance from exoplanet center\\
$z_{\rm ref}$ & Reference $z$-value at one bar \\
$z_{\rm top}$ & $z$-value at the ``top'' of the atmosphere\\
$\beta$ & Angle between ray and local horizon\\
$\kappa$ & Ray curvature\\
$\lambda$ & Wavelength \\
$\mu$ & Atmospheric mean molecular mass \\
$\nu$ & Atmospheric refractivity\\
$\nu_{\rm ref}$ & Reference atmospheric refractivity at one bar\\
$\nu_{\rm STP}$ & STP refractivity\\
$\nu_{\tau=1}$  & Refractivity of a ray with $\tau=1$ \\
$\xi$ & Bending (refraction) angle\\
$\xi_{\tau=1} $ & Bending angle of a ray with $\tau=1$\\
$\sigma_{\rm R}$ & Rayleigh scattering cross section\\
$\tau$ & Path-integrated optical depth \\
$\tau_{\rm R}$ & Rayleigh scattering optical depth \\
$\phi$ & Planetocentric latitude\\
$\chi$ & Rotation angle of ray tracing plane\\
\enddata
\end{deluxetable}

\begin{figure}
\centering
\includegraphics[width=\columnwidth]{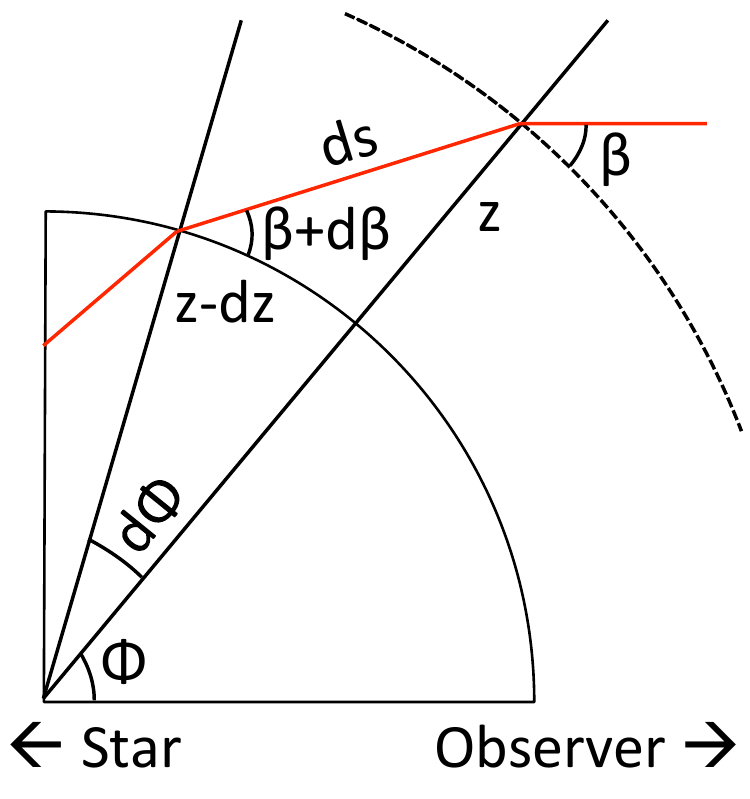}
\caption{Diagram showing a refracted ray path (red) and its parameters within a planetary atmosphere \citep[see also Fig. 1 of ][]{vanderwerf2008}. The concentric semicircles (dashed and solid lines) represent interfaces between regions of different refractive index. \retro\ conducts backwards ray tracing from the observer to the star. By convention, the $\beta$-values in this diagram are negative.}
\label{fig:angles}
\end{figure}

The following expressions determine the path of a ray in a planetary atmosphere \citep{Kivalov2007,vanderwerf2008}.

\begin{align}
\frac{\mathrm{d}z}{\mathrm{d}s} & = \sin{\beta} \label{eq:dz_ds}\\
\frac{\mathrm{d}\beta}{\mathrm{d}s} & = \frac{\cos{\beta}}{z} + \kappa \label{eq:dbeta_ds}\\
\frac{\mathrm{d}\phi}{\mathrm{d}s} & = \frac{\cos{\beta}}{z} \label{eq:dphi_ds}
\end{align}

\noindent where $z$ is the radial distance between the center of the exoplanet and the ray, $\beta$ is the horizon angle swept out between the ray and the local horizon, $\phi$ is the planetocentric latitude with respect to the equator on the observer-side of the atmosphere, and $s$ is the distance along the ray path (Fig. \ref{fig:angles}). In Eq. \ref{eq:dbeta_ds}, $\kappa$ is the local ray curvature defined as 

\begin{equation}\label{eq:curvature}
\kappa = \frac{1}{n} \left [ \cos{\beta} \frac{\mathrm{d} n}{\mathrm{d} z} - \frac{\sin{\beta}}{z} \frac{\mathrm{d} n}{\mathrm{d} \phi} \right ] \;,
\end{equation}
 
\noindent where $n$ is the index of refraction \citep{vanderwerf2008}. The refractivity $\nu$ is related to $n$ by $\nu = n -1$. I simplify the ray curvature by neglecting latitudinal variations in the refractive index (i.e., $\mathrm{d} n / \mathrm{d} \phi = 0$) and assuming spherical symmetry within all exoplanetary atmospheres.

Under the assumption of an isothermal atmosphere in hydrostatic equilibrium, the refractivity $\nu$ follows the atmospheric pressure (and particle number density) according to 

\begin{equation}
\nu(z) = \nu_{\rm ref} \; e^{(z_{\rm ref}-z)/H}
\end{equation}

\noindent where $\nu_{\rm ref}$ is the refractivity at $z_{\rm ref}$, a reference distance from the exoplanet center---typically anchored at one bar of pressure. $H$ is the atmospheric pressure scale height, which satisfies $H=k_B T_{\rm atm}/(\mu g)$ where $k_B$ is the Boltzmann constant, $T_{\rm atm}$ is the atmospheric temperature, $\mu$ is the atmospheric mean molecular mass, and $g$ is the gravitational acceleration.

In addition to $z$, $\phi$, and $\beta$, the bending or refraction angle ($\xi$) is also integrated along the ray path:

\begin{equation}
\frac{\mathrm{d}\xi}{\mathrm{d}s} = \kappa  \label{eq:dxi_ds} \;.
\end{equation}

\noindent The final integrated parameter is the optical depth along the ray path ($\tau$). The specific treatment of optical depth is discussed in \S\ref{sec:param_space}. 

The equations of ray motion are integrated using a fourth-order Runge-Kutta scheme \citep{Runge1895}, which provides a useful balance between computational efficiency and accuracy.

\subsection{Simulating the Stellar Mirage}

To simulate the stellar mirage in the exoplanet atmosphere, I fix the position of the exoplanet at a point in its orbit and trace rays backward from the observer into the equatorial plane of the exoplanet atmosphere. The rays are originally traveling parallel to the line of sight of the distant observer but they diverge due to refraction. As the rays exit the exoplanet atmosphere, the final values of the integrated parameters ($z$, $\beta$, and $\phi$) allow for the projection of the ray paths back to the host star (assuming an index of refraction $n=1$ for interplanetary space). 

For a swath of rays initiated over a range of altitudes in the exoplanet atmosphere, some will impact the star's surface and others will not. The former provide numerical relations between all ray properties including their points of origin on the star, their attenuation, their total bending angles, and their impact parameters with respect to the center of the exoplanet. The latter determine which rays effectively ``bracket'' the star (see the blue symbols in Fig. \ref{fig:crescent_geom}). These rays define the boundary of the secondary image of the star in the exoplanet atmosphere.

A full treatment of the transit geometry requires ray tracing in three dimensions, not just in the equatorial plane. However, under the assumption of a spherically symmetric refractivity profile, three-dimensional ray tracing can be achieved by rotating the plane in which rays are traced through an angle $\chi$ about the center of the exoplanet (Fig. \ref{fig:crescent_geom}). 

Once rotated, the new bounding rays (pink symbols in the bottom panel of Fig. \ref{fig:crescent_geom}) are determined. Now, the plane in which the rays are traced cuts through a chord of the host star instead of the equatorial plane. The entire mirage can be mapped by integrating $\chi$ in the range $\pm \sin^{-1}(R_{\star}/X)$, where $R_{\star}$ is the stellar radius and $X$ is the projected separation between the centers of star and the planet. When the projected disks of the star and planet are mutually tangent, $X$ equals $(R_{\star}+z_{\rm top})$, where $z_{\rm top}$ is the radial distance between the center of the planet and the designated ``top'' of the atmosphere. 

\begin{figure}
\centering
\includegraphics[width=\columnwidth]{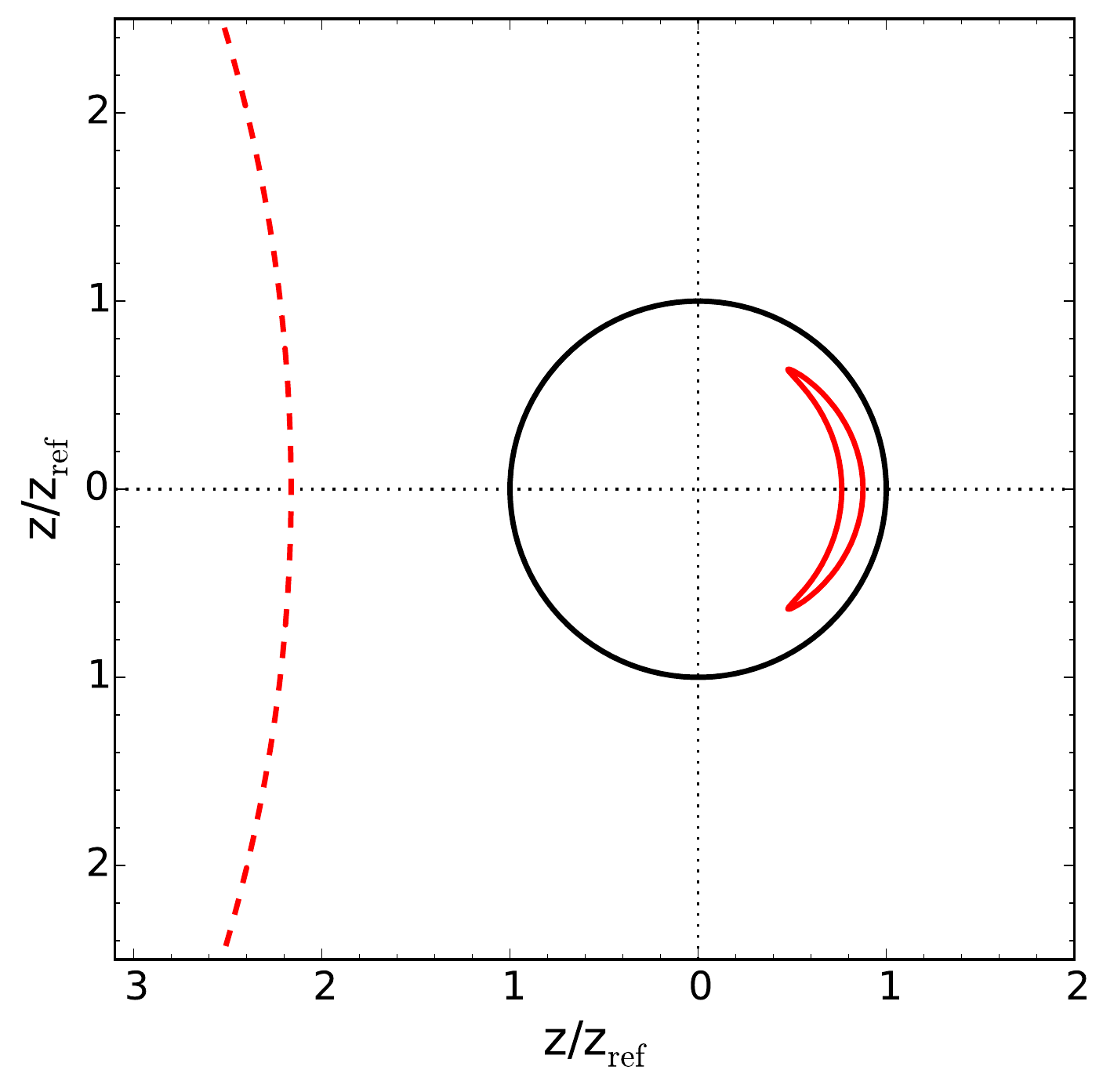}
\caption{Example of the crescent-shaped stellar mirage created by \retro\ for a clear \h\ atmosphere as projected on the sky. The outlines of the star (red, dashed), the one-bar radius of the exoplanet (black),  and the stellar mirage (red, solid) are to scale. In this illustrative---although not strictly physical---case, $R_{\star} = 0.08 R_{\sun}$, the semi-major axis $a = 0.05$ AU, $z_{\rm ref} = 1 R_{\earth}$, $z_{\rm ref}/H$ = 17.2, and the relative flux increase is $\sim$500 parts-per-million. This stellar mirage appears deep within the atmosphere below the one-bar pressure radius. Realistically, the light constituting this mirage would be fully attenuated and a flux increase would not be observed.}
\label{fig:crescent_ex}
\end{figure}

This method yields inverted crescent-shaped images of the host star in the exoplanet's atmosphere (e.g., Fig. \ref{fig:crescent_ex}). The area of the stellar mirage ($A_{\mathcal{M}}$) satisfies the closed line integral

\begin{equation}\label{eq:crescent_area}
A_{\mathcal{M}} = \oint_{\mathcal{M}} \frac{z^2}{2} \,d\chi
\end{equation}

\noindent where $\mathcal{M}$ is the closed curve of the crescent and $z$ is the radial distance as defined before but for points along $\mathcal{M}$. Finally, the relative flux increase ($f_{\mathcal{M}}$) resulting from the appearance of the mirage is simply $f_{\mathcal{M}}=A_{\mathcal{M}}/(\pi \; R_{\star}^2)$. 

All of the steps listed in this section can be repeated as the exoplanet progresses through its near-transit orbit in order to construct a full light curve containing the out-of-transit flux increase. 

\section{Modelling Refracted Light Across Parameter Space}\label{sec:param_space}

The amount of star light that a planetary atmosphere refracts into a distant observer's line of sight is a nonlinear function of stellar, orbital, planetary, and atmospheric parameters.  To identify the regions of parameter space where this phenomenon is detectable in photometric observations, I model the stellar mirage for a comprehensive set of exoplanetary systems. At first, flux attenuation caused by absorption and scattering within the planetary atmosphere is ignored and all atmospheres are assumed to be cloud-free. Therefore, each relative flux increase $f_{\mathcal{M}}$ value is an upper limit. The primary detectability metric is the maximum relative flux increase ($f_{\rm \mathcal{M},max}$) that occurs just before or after transit. This single flux value does not strictly determine whether or not the phenomenon is observable. However, $f_{\rm \mathcal{M},max}$ is sufficient to identify favorable regions of parameter space.

I model exoplanetary systems with various values of these five parameters (Table \ref{tab:params}): 
\begin{itemize}
\item \textit{Stellar radius ($R_{\star}$)}. The size of the host star is related to the size of the secondary image in the planetary atmosphere. F, G, K, and M dwarf stars down to the canonical hydrogen burning limit are considered. 
\item \textit{Orbital semi-major axis} ($a$). The orbital distance is related to the amount of bending a ray must experience to enter the line of sight of the observer. Circular orbits are used throughout this work, so $a$ is equivalent to orbital distance at all points in the orbit.\footnote{Hereafter, the terms ``orbital distance'' and ``semi-major axis'' will be used interchangeably. This is only true for orbits with eccentricities of zero, which are used throughout this work.} Orbital eccentricity influences the morphology of the relative flux increase, but does not alter its maximum value. Short- and long-period orbits are considered to capture hot-Jupiters and Saturn-analogs alike.
\item \textit{Planet mass} ($M_{\rm p}$). Planet mass influences the size of a planet and its atmospheric structure. Considering Earth-mass to Jovian-mass planets, $M_{\rm p}$ determines the planetary radius ($z_{\rm ref}$) via the empirical, deterministic mass-radius relations of \citet{Chen2017}. The radius returned from this mass-radius relation is treated as the one-bar reference radius, which anchors the atmospheric refractivity profile. The planet mass and radius yield the gravitational acceleration ($g$), which is assumed to be constant throughout the atmosphere. The planet radius is arguably more fundamentally related to atmospheric refraction. However, for the planetary mass and radius ranges under consideration, empirical relations suggest that mass is not a single-valued function of radius \citep[e.g.,][]{Chen2017}. Increasing the planetary mass eventually leads to gravitational self-compression such that more massive planets are smaller. This process influences the atmospheric scale height, which plays a role in determining the size of the stellar mirage.
\item \textit{Atmospheric temperature} ($T_{\rm atm}$). The extent of the atmosphere---as dictated by the scale height $H$---is proportional to $T_{\rm atm}$. This temperature may be correlated with semi-major axis, so it is possible to assume $T_{\rm atm}$ equals the planetary equilibrium temperature. However, a better approach to exploring the relation between temperature and the stellar mirage is to decouple $T_{\rm atm}$ and $a$ and allow any planet at any semi-major axis to have any temperature. This treatment incorporates more exotic scenarios such as young, long-period gaseous exoplanets that are still hot from gravitational contraction and exoplanets experiencing the greenhouse effect. All atmospheres are assumed to be isothermal. If the altitude range sampled by the rays that create the stellar mirage is small, then this assumption is valid. Atmospheric temperatures between 50 and 1000 K are considered. 
\item \textit{Atmospheric composition}. The chemical composition of the atmosphere influences its mean molecular mass ($\mu$) and refractivity ($\nu$). I use refractivity values derived from laboratory measurements of various gases at 101325 Pa and 273.15 K, or ``Standard Temperature and Pressure'' (STP) at the time the measurements were made \citep[][Section 2.5.7]{Kaye1995}. These refractivities have the symbol $\nu_{\rm STP}$. Assuming the ideal equation of state for the atmospheric gas, $T_{\rm atm}$ gives the corresponding one-bar (1\e{5} Pa) refractivity values ($\nu_{\rm ref}$). Atmospheric composition itself is a vast parameter space, so I choose the end-member cases \h, 100\% \water, 100\% \n, 100\% \carbon\ to obtain a general sense of how the chemical composition influences the stellar mirage. In the \h\ case, the helium mass fraction is given the solar value of $Y=0.25$ \citep{Asplund2009}. Table \ref{tab:munu} provides values of $\mu$ and $\nu_{\rm STP}$.
\end{itemize}

\begin{deluxetable}{lc}
\tablecaption{Definition of the Rectilinear Parameter Space Grid\label{tab:params}}
\tablecolumns{2}
\tablewidth{\columnwidth}
\tablehead{
\colhead{Parameter} &
\colhead{Domain}
}
\startdata
Stellar radius $R_{\star}$ ($R_{\sun}$) & [0.08,1.5]\tablenotemark{a}  \\
Orbital semi-major axis $a$ (AU) & [0.05,10]\tablenotemark{b} \\
Planet mass $M_{\rm p}$ ($M_{\earth}$) & [1,400]\tablenotemark{b} \\
Atmospheric temperature $T_{\rm atm}$ (K) & [50,1000]\tablenotemark{b} \\
Atmosphere type & \h, \water, \n, \carbon \\
\enddata
\tablenotetext{a}{Values are evenly sampled in linear space.}
\tablenotetext{b}{Values are evenly sampled in logarithmic space.}
\end{deluxetable}

\begin{figure}
\centering
\includegraphics[width=\columnwidth]{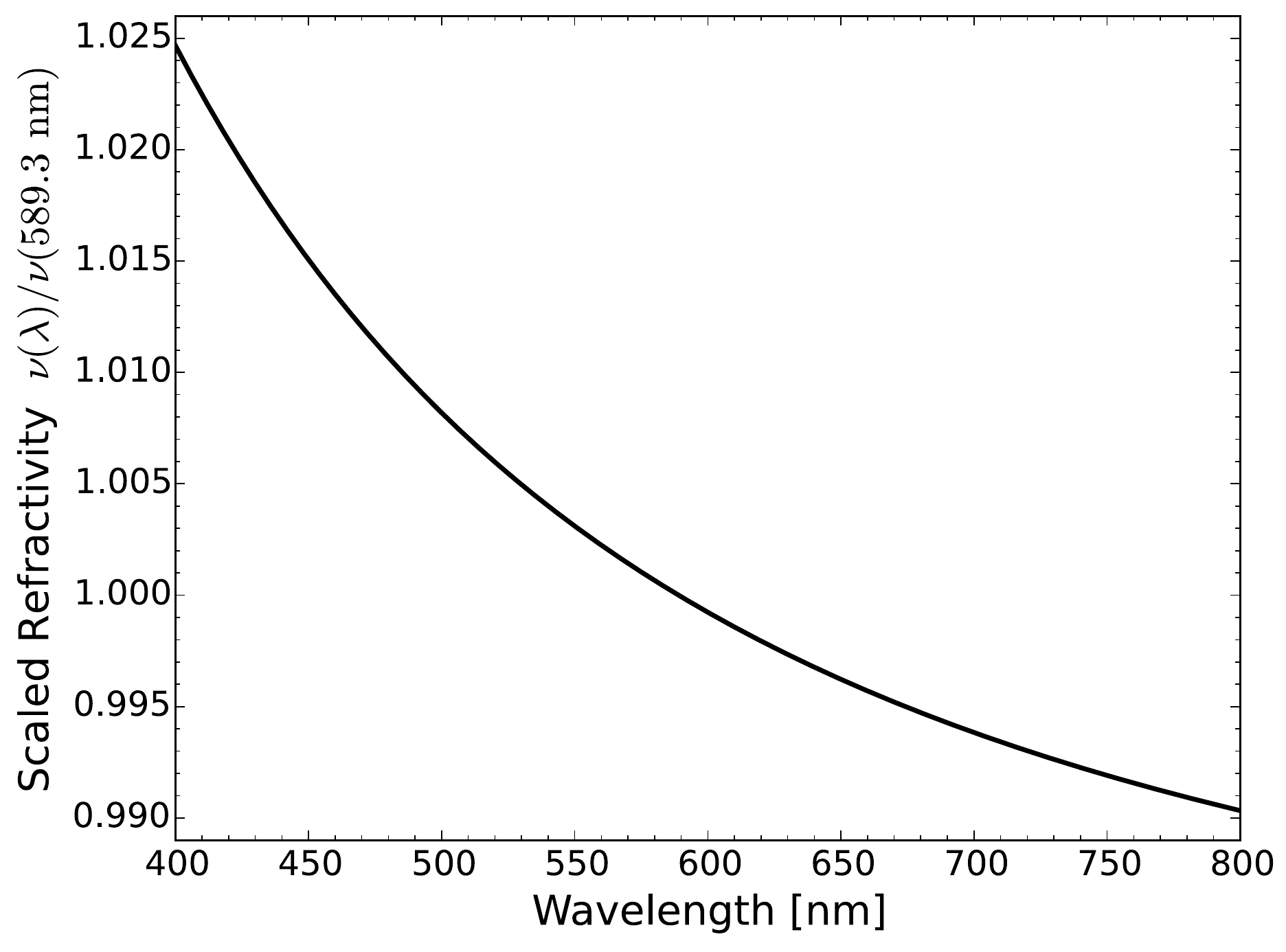}
\caption{Jupiter's atmospheric refractivity at visible wavelengths relative to 589.3 nm assuming a helium mole fraction of 0.14 \citep{vonZahn1998} and refraction and dispersion coefficients from \citet[][pp. 101]{Born1999}.}
\label{fig:ior}
\end{figure}

Refractivity is a wavelength-dependent quantity (e.g., Fig. \ref{fig:ior}). Away from electronic, vibrational, and rotational transitions, the wavelength dependence of $\nu$ is approximated by Cauchy's formula: $\nu = \mathcal{A}(1+\mathcal{B}/\lambda^2)$ where $\lambda$ is wavelength and $\mathcal{A}$ and $\mathcal{B}$ are the coefficients of refraction and dispersion of the medium, respectively. The STP refractivity values in Table \ref{tab:munu} are measured at 589.3 nm, so all modeling hereafter is quantitatively valid at that wavelength. However, the results of this work are qualitatively applicable across much of the visible portion of the electromagnetic spectrum. Jupiter's atmospheric refractivity, for example, only changes by $\sim$3.5\% from the blue end to the red end of the visible regime (Fig. \ref{fig:ior}). I return to the wavelength-dependent nature of refractivity in \S\ref{sec:wavelength}.

The atmospheric scale height is not directly varied, as it is a function of several fundamental parameters. Altering the planet mass (and therefore the radius and gravity), the atmospheric temperature, and the atmospheric composition ensures a wide variety of $H$-values is considered.

Each parameter is sampled 12 times within its bounds, with the exception of atmospheric composition that only has four values. The result is a rectilinear grid of 4$\times$12$^4 = 82944$ nodes defining the full parameter space. The stellar mirage as a function of orbital phase leading up to transit is modeled at each of these nodes. The result is 82944 transit light curves displaying the ``shoulders'' caused by refracted light. Atmospheric variations between the dusk and dawn portions of the planetary atmosphere are not considered, so the light curve before transit ingress is identical to the one after transit egress.

Some of the grid points represent exotic or even unphysical planetary systems. For the sake of completeness, I model those cases to fill the grid and understand where atmospheric refraction could potentially create observable stellar mirages. 

\begin{deluxetable}{lcc}
\tablecaption{Mean Molecular Mass and STP Refractivity\label{tab:munu}}
\tablewidth{\columnwidth}
\tablehead{
\colhead{Composition} &
\colhead{$\mu$ ($m_{\rm H}$)} &
\colhead{$\nu_{\rm STP}$\tablenotemark{a}}
}
\startdata
\h & 2.29 & 1.18\e{-4} \\
100\% \water & 18 & 2.56\e{-4} \\
100\% \n & 28 & 2.98\e{-4}\\
100\% \carbon & 44 & 4.49\e{-4}\\
\enddata
\tablenotetext{a}{These refractivity values were measured at 101325 Pa and 273.15 K \citep[][Section 2.5.7]{Kaye1995}. The \h\ value is a combination of $\nu_{\rm STP}=$ 1.32\e{-4} for H$_2$ and $\nu_{\rm STP}=$ 3.5\e{-5} for He at the mole fraction corresponding to $Y=0.25$ (solar).}
\end{deluxetable}

\subsection{Opacity}

For \h\ atmospheres, nominal sources of opacity include (but are not limited to) Rayleigh scattering, H$_2$-He collision-induced absorption (CIA), and absorption from trace species such as CH$_4$ and H$_2$O. It is well established that absorption, not refraction, will dictate the transmission of flux through the atmospheres of short-period exoplanets \citep[e.g.,][]{Seager2000,Hubbard2001}. At long-periods, atmospheres are typically cold and H$_2$O is likely to be cold-trapped into a cloud layer that is difficult to remotely sense. This is indeed the case for Jupiter and Saturn. Therefore, it is reasonable to consider CH$_4$ as the primary absorbing trace species for the atmosphere modeling. At 589.3 nm, the extinction coefficient of a 1\% mixing ratio of CH$_4$ in the \h\ atmosphere at STP is $\sim$2.7\e{-8} m$^{-1}$ \citep{Burrows2001}.\footnote{For perspective, the canonical atmospheric mixing ratios of CH$_4$ for Jupiter and Saturn are 0.3\% and 0.45\%, respectively.} However, the extinction coefficient from Rayleigh scattering in the same atmosphere is $\sim$1.4\e{-6} m$^{-1}$. The contribution of CH$_4$ to the total extinction is $\lesssim$2\%. Similarly, the 589.3 nm extinction coefficient for H$_2$-He CIA at STP in the \h\ atmosphere is $\sim$1.2\e{-11} m$^{-1}$ \citep{Richard2012}, which is also negligible compared to the Rayleigh scattering extinction. Clearly, the primary source of opacity for the models in this work is Rayleigh scattering.

The optical depth due to Rayleigh scattering along the ray path ($\tau_{\rm R}$) satisfies

\begin{equation}
\tau_{\rm R} = \int \; \sigma_{\rm R}(z) \; N(z) \, ds   
\end{equation}

\noindent where $N$ is the total number density and $\sigma_{\rm R}$ is the Rayleigh scattering cross section. Following \citet[][pp. 167]{Seager2010},

\begin{equation}
\sigma_{\rm R}(z) = \frac{24 \pi^3}{N(z)^2 \lambda^4} \left [\frac{n(z)^2-1}{n(z)^2+2}\right ]^2 \;.
\end{equation}

Cloud layers can present a significant source of opacity to the stellar mirage \citep{Misra2014b}.  Clouds potentially truncate the refracted light signal above some pressure level (below some altitude). In the case of a Saturn-analog exoplanet, the atmospheric ``surface'' created by refraction would exist above the cloud deck in altitude \citep{Dalba2015}. Here, I do not consider opacity from clouds.

Sources of opacity in the other atmosphere types (i.e., \water, \n, and \carbon) are not initially included to test whether or not the stellar mirage is even detectable in the best case of a clear atmosphere.

\section{Results}\label{sec:results}

\begin{figure*}
\centering
\includegraphics[width=\textwidth]{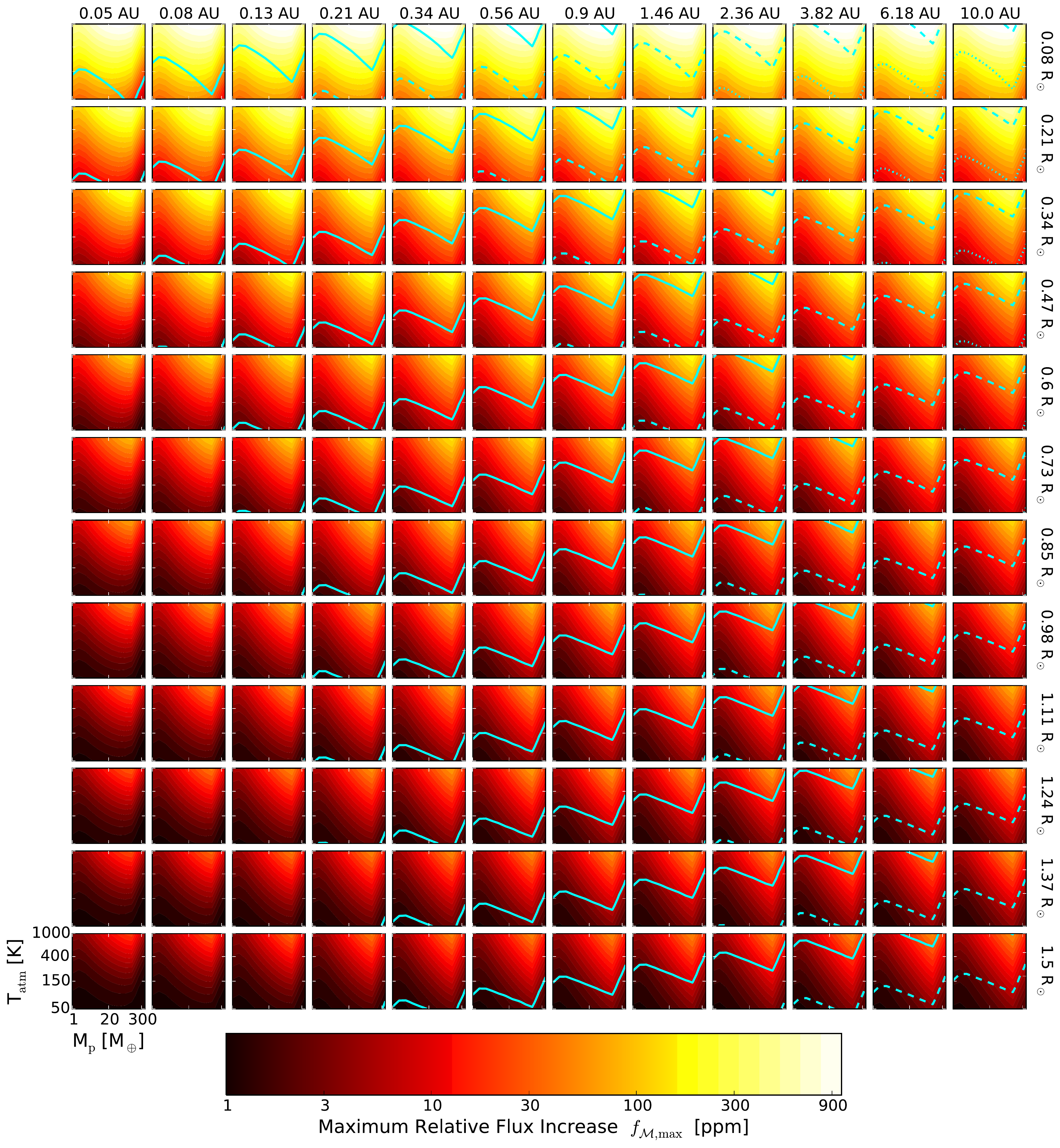}
\caption{Maximum relative flux increase ($f_{\rm \mathcal{M},max}$) due to the appearance of stellar mirages in \h\ atmospheres of various exoplanet systems. The axes and scales of each individual panel are identical to those on the bottom left panel. Each column of panels is evaluated at a single value of semi-major axis $a$, and each row of panels is evaluated at a single value of stellar radius $R_{\star}$. The blue solid, dashed, and dotted lines represent Rayleigh scattering optical depths ($\tau_{\rm R}$) of 10, 1, and 0.1, respectively. Any panel without a blue line is entirely greater than $\tau_{\rm R}=10$. At visible wavelengths, the detectability of stellar mirages is limited to long-period, Saturn-mass planets orbiting M dwarf stars and perhaps any cold planet orbiting an ultra-cool dwarf star beyond $\sim$0.5 AU.}
\label{fig:contour_full}
\end{figure*}

The primary results of the transiting exoplanet parameter space exploration are shown in Fig. \ref{fig:contour_full}, which displays variation within four dimensions of the total parameter space. The heat map colors correspond to the maximum relative increase in flux ($f_{\rm \mathcal{M},max}$) caused by atmospheric refraction. This maximum increase occurs at the moment prior to (or after) transit, and is found assuming a clear atmosphere (i.e., no absorption or scattering). All systems displayed in Fig. \ref{fig:contour_full} have \h\ atmospheres. Each individual panel displays $f_{\rm \mathcal{M},max}$ for 144 combinations of atmospheric temperature $T_{\rm atm}$ and planet mass $M_{p}$. Their scales are displayed on the bottom left panel, and are identical for all other panels. Each $f_{\rm \mathcal{M},max}$ value in a single panel has a single value of semi-major axis $a$ and stellar radius $R_{\star}$, which are displayed as ``secondary'' x and y axes on the top and at the right of the figure. Similar visualizations exist for the other three atmospheric compositions (\water, \n, and \carbon), but the relative flux increase values are significantly smaller so they are not displayed. Hereafter, I only consider cases of \h\ atmospheres because refraction phenomena are unlikely to be observed in other atmosphere types until photometry with single part-per-million (ppm) precision is readily achievable. 

The blue solid, dashed, and dotted lines in Fig. \ref{fig:contour_full} represent Rayleigh scattering optical depth $\tau_{\rm R}$ values of 10, 1, and 0.1, respectively for the equatorial ray at the near edge of the stellar mirage (nearest to the exoplanet) just before or after transit. In reality, there is a slight decrease in $\tau_{\rm R}$ between the near and far edges of the mirage. The inner equatorial ray is a conservative benchmark to understand where the mirage starts to become attenuated. If the inner equatorial ray has $\tau_{\rm R}<1$, then the entire stellar mirage does as well. As the projected planet-star separation increases, however, the stellar mirage appears deeper in the atmosphere. As a result, for exoplanet systems with $\tau_{\rm R} \approx 1$, the refracted light signal may not be optically thin until the exoplanet is sufficiently near to transit. 

Three trends in the appearance of stellar mirages in \h\ atmospheres are apparent in Fig. \ref{fig:contour_full}. First, refracted light is not significant for short-orbit exoplanet systems of any kind because of the dominating opacity from Rayleigh scattering. As discussed in \S\ref{sec:intro}, this result is expected \citep[e.g.,][]{Seager2000,Hubbard2001}. Below $\sim$0.2 AU, no exoplanet systems within the parameter space in consideration display an optically thin refracted light signal. Up to $\sim$6 AU, some optically thin pockets of parameter space exist where the maximum relative flux increase can be on the order of tens of ppm. These favorable pockets typically occur for stars with sub-solar radii. By 10 AU, most gas giant exoplanets transiting stars with radii $R_{\star} \lesssim R_{\sun}$ create clear stellar mirages. These approximate limits are in partial disagreement with \citet{Sidis2010}, who claim that refraction effects are significant for exoplanets on orbits $\gtrsim$70 days---which corresponds to $\gtrsim$0.33 AU for a Sun-like star. This discrepancy is potentially a result of the toy model employed by \citet{Sidis2010} to describe the shape and size of the stellar mirage.

Fig. \ref{fig:contour_full} demonstrates that any individual parameter (e.g., orbital period) cannot adequately describe whether or not refraction effects will be significant for a given planetary system. At the very least, the value of $a/R_{\star}$ is indicative of the size of the stellar mirage. In practice, the appearance of the out-of-transit stellar mirage \emph{at visible wavelengths} should be prevalent (i.e., maximum relative flux increase values of at least tens of ppm) for exoplanets with \h\ atmospheres and $a/R_{\star} \gtrsim$ 4000. This region occupies the top-right corner of Fig. \ref{fig:contour_full}. Within this range, the atmospheric temperature $T_{\rm atm}$ in part determines whether Rayleigh scattering or refraction is the dominant process. A realistic estimate for $T_{\rm atm}$ is the planetary equilibrium temperature defined as 

\begin{equation}
T_{\rm eq} = T_{\rm eff}(1-A_{\rm B})^{1/4}\sqrt{\frac{R_{\star}}{2a}}
\end{equation}

\noindent where $T_{\rm eff}$ is the stellar effective temperature, $A_{\rm B}$ is the planetary Bond albedo, and perfect energy redistribution is assumed. For the conservative values of $a/R_{\star} \gtrsim 4000$, $A_{\rm B}=0$, and $T_{\rm eff}<4000$ K, the equilibrium temperature is $T_{\rm eq} \lesssim $ 45 K. The x-axis of each panel in Fig. \ref{fig:contour_full} rests at $T_{\rm atm}=50$ K, well into the clear atmosphere regime below an optical depth of unity. Therefore, cold exoplanets satisfying the $a/R_{\star} \gtrsim 4000$ criterion will display observable stellar mirages at visible wavelengths.    

The second trend in Fig. \ref{fig:contour_full} is that the magnitude of $f_{\rm \mathcal{M},max}$ increases with the atmospheric scale height $H$. This trend is evident in the $f_{\rm \mathcal{M},max}$ variations as functions of $T_{\rm atm}$ and mean molecular mass. Increasing $T_{\rm atm}$ leads to greater maximum relative flux increases, and the high mean molecular mass atmospheres (not shown) display correspondingly smaller flux signals. This trend is in agreement with \citet{Sidis2010}.

The third trend in Fig. \ref{fig:contour_full} is the increase in $f_{\rm \mathcal{M},max}$ values in cold atmospheres for Saturn-mass ($\sim$100-$M_{\earth}$) exoplanets. This feature is a direct result of the gravitational self-compression that is present in the empirical mass-radius relations from \citet{Chen2017}. For Saturn-mass planets, the local minimum in gravitational acceleration creates larger values of atmospheric scale height. Also, the large planetary radius values create long ray path lengths and considerable bending angles. The opacity displays a similar feature as it is also a path-integrated quantity. The increased signal for Saturn-mass exoplanets suggests that low density or inflated exoplanets are quite favorable for refraction related phenomena. To date, all such exoplanets exist on relatively short-period orbits and are unlikely to display any effects of refracted light. 

There is one pocket of parameter space that produces observable stellar mirages but does not satisfy the $a/R_{\star} \gtrsim 4000$ criterion. Exoplanets of \emph{any} mass orbiting beyond $\sim$0.5 AU around late M or ultra-cool dwarf stars with radii $R_{\star} \lesssim 0.1R_{\sun}$ have maximum relative flux increase values greater than 50 ppm. For a late M dwarf star with $T_{\rm eff} = 2500$ K, the corresponding equilibrium temperature of an exoplanet at 0.5 AU with $A_{\rm B}=0$ is $\sim$54 K, which ensures the optical depth is less than unity for almost any value of planet mass. Although an interesting case, cool stars like this are faint at visible wavelengths and achieving even 100 ppm precision would be a formidable observational challenge.

\section{Atmospheric Lensing by Non-Transiting Exoplanets}\label{sec:lens}

\begin{figure}
\centering
\includegraphics[width=\columnwidth]{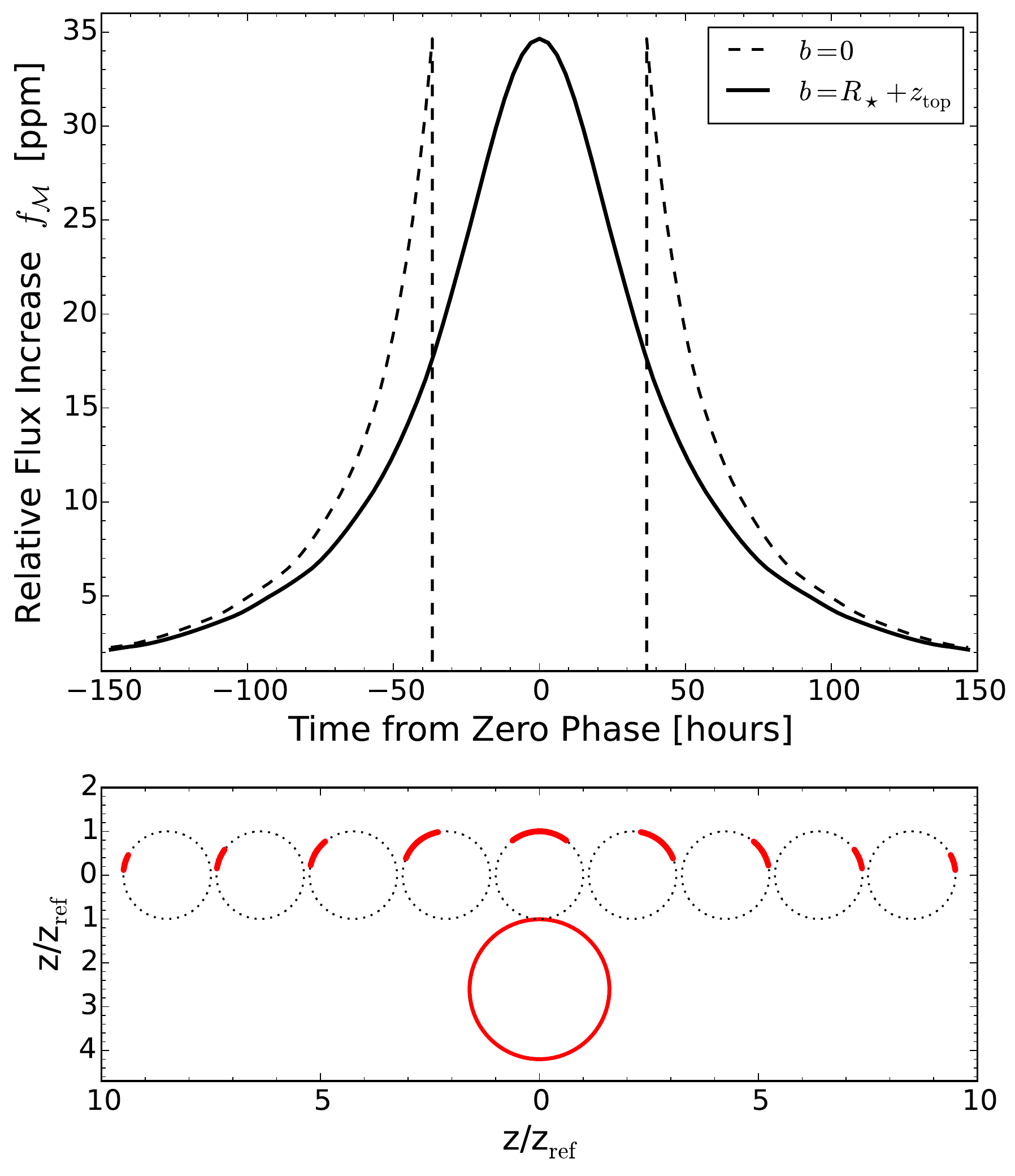}
\caption{\retro\ model of a system with the following parameters: Clear \h\ atmosphere, $a=$ 1.46 AU, $R_{\star}=$ 0.21$R_{\sun}$, $T_{\rm atm}=T_{\rm eq}=$ 50 K, $M_{\rm p}=$ 1.4$M_{\rm Saturn}$, and $z_{\rm ref}=$ 1.6$R_{\rm Saturn}$. {\bf Top:} The top panel shows the relative flux increase caused by refraction for transiting (dashed) and non-transiting (solid) orbits. For the transiting case, $f_{\mathcal{M}}$ values decrease below zero as refraction causes an overall decrease in flux during transit. The atmospheric lensing event that occurs in the non-transiting case is potentially more detectable than the ``shoulders'' from the transiting case. {\bf Bottom:} The bottom panel shows the projected positions of the exoplanet (dotted), host star (red circle), and stellar mirages (red arcs) for a non-transiting orbit to scale. The stellar mirages appear above z$_{\rm ref}$ in altitude and have finite widths and areas. The mirage advances in latitude as the exoplanet orbits the star, reaching its maximum size at an orbital phase of zero.}
\label{fig:lens-crescent_rot}
\end{figure}

The amount of starlight refracted into an observer's line of sight by an exoplanetary atmosphere is maximal when the projected disks of the planet and star are mutually tangent. As the projected separation ($X$) increases, the physical extent of the stellar mirage diminishes along with the relative flux increase $f_{\mathcal{M}}$. I have, so far, only considered transiting exoplanets with edge-on orbits. In this case, the transit impact parameter ($b$) is zero, and the exoplanet's projected motion is either directly toward or away from the center of the host star. As a result, the rate at which $f_{\mathcal{M}}$ increases (or decreases) is also maximal. If the projected planet-star separation shrinks more slowly, then $f_{\mathcal{M}}$ changes more gradually. With this in mind, the greatest out-of-transit refracted light signal is achieved by a \emph{non-transiting} exoplanet with impact parameter $b=R_{\star}+z_{\rm top}$ (Fig. \ref{fig:lens-crescent_rot}). 

Figure \ref{fig:lens-crescent_rot} contains the modeled light curves of an exoplanet system considering both a transiting ($b$=0) and non-transiting ($b=R_{\star}+z_{\rm top}$) orbit. The transiting case produces ``shoulders''---sharp increases in flux---that sit on the edges of the transit light curve. The non-transiting case creates an ``atmospheric lensing event'' as the stellar mirage slides along exoplanet's terminator. In both cases, the maximum relative flux increase is the same because the same minimum projected planet-star separation is achieved. However, the atmospheric lensing event that occurs in the non-transiting case is several times longer than the relatively short flux increases on either side of transit. The full width at half maximum of the lensing feature is approximately equivalent to the length of the transit that would occur for an edge on transit. Both types of lensing events shown in Fig. \ref{fig:lens-crescent_rot} are significantly longer than transits of presently known exoplanets owing to the low mass and long-period of the modeled exoplanet ($a=$ 1.46 AU, $R_{\star}=$ 0.21$R_{\sun}$, $T_{\rm atm}=T_{\rm eq}=$ 50 K, $M_{\rm p}=$ 1.4$M_{\rm Saturn}$, and $z_{\rm ref}=$ 1.6$R_{\rm Saturn}$). However, this particular system is one of the few that produces potentially observable signals that overcome opacity from Rayleigh scattering at visible wavelengths. 

The detection of exoplanets via the transit method is intrinsically biased against those with long-periods. Assuming a circular orbit, the simple geometric transit probability is $P_{\rm transit} \approx R_{\star}/a$. The chances of detection are slightly improved for an exoplanet with an atmosphere prone to refraction. Including refraction effects, the detection probability is $P_{\rm detect} \approx P_{\rm transit} + z_{\rm top}/a$. Since the out-of-transit refracted light signal is quite significant for many cases of gas giants orbiting small stars, refraction can theoretically almost double the geometric probability of detecting a long-period exoplanet.

\subsection{Detectability of Atmospheric Lensing in the \kepler\ Data Set}\label{sec:detect}

I investigate the likelihood of a refracted light signal from a non-transiting exoplanet existing in the \kepler\ data set. Instead of basing the investigation on systems with currently known exoplanets, I focus on the precision of the \kepler\ photometry. A consideration of refracted light signals from known exoplanetary systems is provided in \S\ref{sec:disc}.

Stellar flux that is collected by a particular telescope and instrument creates a rate of measured photoelectrons per unit time (denoted by the symbol $J$). Considering only Poisson noise, the total signal-to-noise ratio (SNR$_{\rm total}$) of a single atmospheric lensing event is 

\begin{equation}\label{eq:snr}
{\rm SNR}_{\rm total} = \frac{J \int_0^T f_{\mathcal{M}}(t) \; dt}{\sqrt{JT + \int_0^T f_{\mathcal{M}}(t) \; dt}}
\end{equation}

\noindent where $t$ is time and $T$ is the total duration of the atmospheric lensing event. Given $J$, Eq. \ref{eq:snr} returns the maximum achievable SNR for any exoplanet atmospheric lensing event. 

The \kepler\ spacecraft \citep{Borucki2010,Koch2010} launched in 2009 and made high-precision photometric observations of a single patch of sky until the end of its primary mission in 2013. The \kepler\ bandpass spans 430 nm to 880 nm, so the results of this refraction study are applicable to the \kepler\ data set. By specifying $J$ as a function of relative stellar magnitude for the \kepler\ spacecraft,\footnote{Empirical \kepler\ photoelectron rates are available at \url{https://keplergo.arc.nasa.gov/CalibrationSN.shtml}.} I use \retro\ models to calculate SNR$_{\rm total}$ for all 82944 exoplanet systems from \S\ref{sec:param_space}. For all exoplanets, non-transiting orbits with impact parameters $b=R_{\star}+z_{\rm top}$ are assumed, which yield the greatest values of maximum relative flux increase $f_{\rm \mathcal{M},max}$. For $b>R_{\star}+z_{\rm top}$, the duration of the atmospheric lensing event increases but $f_{\rm \mathcal{M},max}$ decreases. In practice, the increase in detectability gained for a longer lensing event is eventually lost due to the low peak-to-baseline contrast.

To be conservative, the denominator in Eq. \ref{eq:snr} is replaced with an informed estimate of the precision achieved by actual \kepler\ data. The 10th percentile precision of 10th magnitude\footnote{\kepler-band magnitude} stars observed by \kepler\ is 12.3 ppm in 6 hours of observation \citep{Christiansen2012}. The SNR$_{\rm total}$ for {\it Kepler's} 10th-magnitude stars is then found by integrating the numerator in Eq. \ref{eq:snr} and dividing by the measured \kepler\ precision scaled to the total duration of the lensing event. The single event SNR$_{\rm total}$ values range from 0 to 250 across the entire parameter space. As discussed previously, though, certain portions of parameter space are unphysical or are subject to significant Rayleigh scattering. The threshold for an actual detection is more complicated than a single SNR estimate, so any distinction made here between ``detectable'' and ``not detectable'' is approximate. 

\begin{figure}
\centering
\includegraphics[width=\columnwidth]{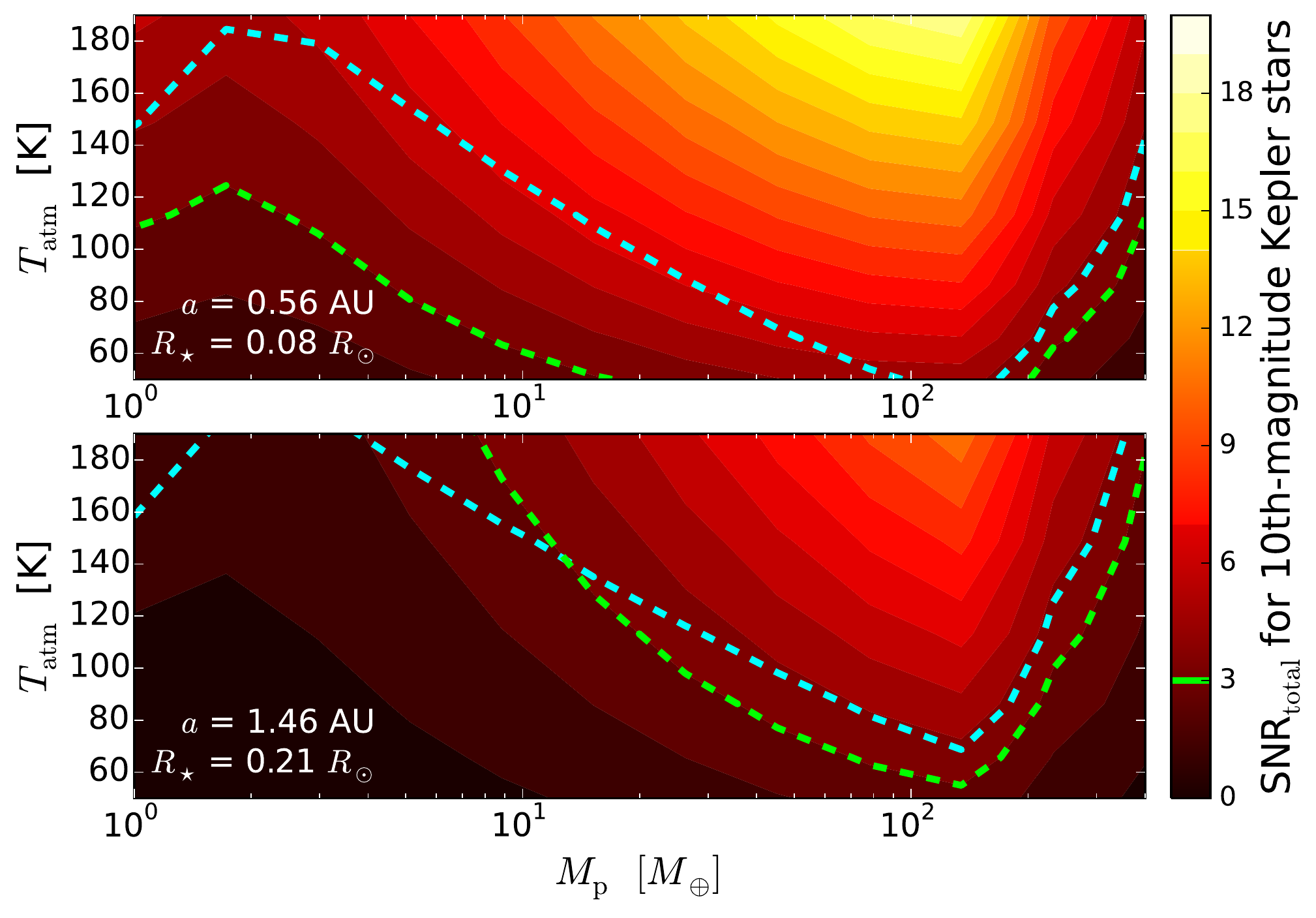}
\caption{Modeled detectability of atmospheric lensing events caused by potential non-transiting exoplanets in \kepler\ data. The SNR$_{\rm total}$ values are found for clear \h\ atmospheres, the displayed parameters, and the measured \kepler\ precision for 10th-magnitude stars \citep{Christiansen2012}. The blue dashed line represents the Rayleigh scattering optical depth of unity decreasing to lower $T_{\rm atm}$ values. The green dashed line identifies a single event SNR$_{\rm total}=3$ for reference. {\bf Top:} Sub-Saturn mass exoplanets orbiting late M or ultra-cool dwarf stars can produce observable signatures of refracted light in \kepler\ data, although finding a suitably bright host star is potentially impractical. {\bf Bottom:} The atmospheres of sub-Jupiter mass exoplanets on 1--2 AU orbits around small stars may also create detectable lensing events in the \kepler\ data set.}
\label{fig:detect}
\end{figure}

The results of the signal-to-noise ratio calculation within two subsets of the entire parameter space are shown in Fig. \ref{fig:detect}. Each panel contains varying atmospheric temperatures ($T_{\rm atm}$) and planet masses ($M_{\rm p}$) for clear \h\ atmospheres and single values of semi-major axis ($a$) and stellar radius ($R_{\star}$). In both panels, the planetary equilibrium temperature for a Bond albedo of zero is approximately 50 K, so realistic exoplanets would lie on or near the x-axes. The top panel shows that SNR$_{\rm total}$ values of 5--10 are achievable for sub-Saturn mass exoplanets orbiting late M or ultra-cool dwarf stars. This portion of parameter space exists below the blue Rayleigh scattering optical depth ($\tau_{\rm R}=1$) line, suggesting that the assumption of a clear atmosphere in calculating the relative flux increase (and SNR$_{\rm total}$) is valid. However, given the faintness of small stars at visible wavelengths, such a detection may be infeasible. The bottom panel of Fig. \ref{fig:detect} highlights a narrow region of detectability for exoplanets with masses between those of Saturn and Jupiter, on $\sim$1.5 AU orbits around slightly larger M dwarf stars. However, the combination of SNR$_{\rm total} \approx 3$ and $\tau_{\rm R} \approx 1$ decreases the likelihood of detection in this region. 

The SNR$_{\rm total}$ values for 10th magnitude \kepler\ stars generally display the same trends discussed in \S\ref{sec:param_space}. Opacity from Rayleigh scattering at visible wavelengths mostly restricts detections of stellar mirages in \kepler\ photometry to either rare or impractical pockets of parameter space (e.g., $a \gtrsim 6$ AU). Additionally, disentangling the atmospheric lensing feature from sources of stellar variability is a challenging endeavor. This is especially difficult for systems that have lensing event durations similar to the stellar rotation period. Moreover, it is difficult to prioritize a search for an atmospheric lensing event. Since atmospheric lensing occurs for non-transiting planets, the signal could exist in almost any light curve, and prioritizing, for instance, the \kepler\ Objects of Interest (KOIs) may be unhelpful. 

In conclusion, atmospheric lensing can create photometric signals that are theoretically detectable within data at \kepler's precision. However, Rayleigh scattering likely attenuates or entirely removes any such signals.

\section{Discussion}\label{sec:disc}

\subsection{For which systems is refraction important?}

Atmospheric retrieval from transmission spectra is a useful technique to characterize exoplanetary atmospheres. Many retrieval studies \citep[e.g.,][]{Benneke2012,Line2012,Barstow2013,Lee2014,Waldmann2015,Morley2017} use the geometric limit where rays travel in straight-line paths, effectively ignoring refraction. As the number of potentially characterizable, long-period transiting exoplanet candidates grows \citep[e.g.,][]{Kipping2014a,Wang2015,Kipping2016b,Uehara2016,ForemanMackey2016,Osborn2016,Morton2016}, it seems useful to estimate at what point refraction should not be ignored.

\citet{Hui2002} made the first attempt to determine for which systems refraction is important. They developed a formalism based on the existence of caustics---positions in the source (host star) plane where lensing caused by atmospheric refraction results in diverging magnification of the host star. In doing so, \citet{Hui2002} defined a useful parameter, $B_{\rm HS02} = 2 a \nu_{\tau=1}/H$ where $\nu_{\tau=1}$ was the atmospheric refractivity probed by a ray that achieved an optical depth of unity, respectively. In a spherically symmetric atmosphere, refractivity relates to bending angle such that $\nu_{\tau=1} = \xi_{\tau=1} \sqrt{H/(2\pi b_{\tau=1})}$ where $\xi_{\tau=1}$ and $b_{\tau=1}$ are the bending angle and impact parameter of the ray that achieves an optical depth of unity. Here, $\xi_{\tau=1}$ is positive by convention. Substituting this relation into $B_{\rm HS02}$ gives 

\begin{equation}\label{eq:B}
B_{\rm HS02} = a \xi_{\tau=1} \sqrt{\frac{2}{\pi H b_{\tau=1}}}
\end{equation}

\noindent For a spherical planet, \citet{Hui2002} found that $B_{\rm HS02}$ and $b_{\tau=1}/H$ (a balance between the atmospheric thermal and binding energies) described the importance of atmospheric lensing. They derived a condition for ``strong lensing,'' where atmospheric refraction was deemed important \citep[Eq. 13 of ][]{Hui2002}:\footnote{I ignore the planetary oblateness when reproducing the relations of \citet{Hui2002}.} 

\begin{equation}\label{eq:strong_lens}
1-\sqrt{\frac{\pi H}{2b_{\tau=1}}}B_{\rm HS02} < 0
\end{equation}

\begin{figure}
\centering
\includegraphics[width=\columnwidth]{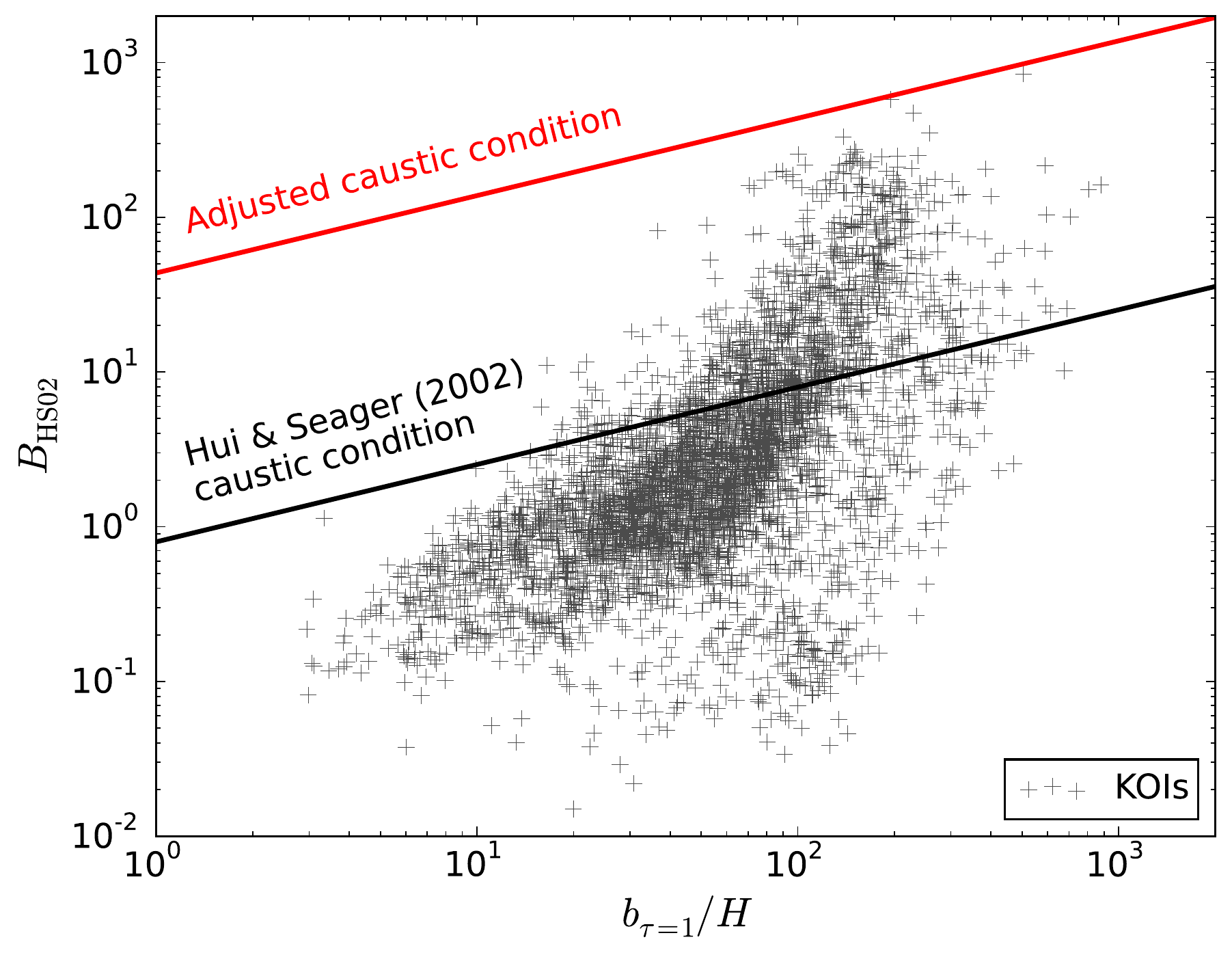}
\caption{Updated version of Fig. 3 from \citet{Hui2002} illustrating the condition for caustics (refraction) in transiting exoplanet systems. The gray markers are KOIs with \h\ atmospheres modeled with \retro\ at visible wavelengths. According to \citet{Hui2002}, any planets above the black line display caustics (i.e., refraction dominates Rayleigh scattering). I suggest (in the text) that this condition is overly optimistic, potentially due to the mathematical formalism employed by \citet{Hui2002}. The red line is the adjusted caustic condition, which is increased by the KOIs' median value of $R_{\star}/b_{\tau=1}$. This adjusted condition is more consistent with the detectability of out-of-transit refracted light presented in \S\ref{sec:param_space} and \S\ref{sec:lens}.}
\label{fig:B}
\end{figure}

This condition broadly divided parameter space into regions with or without caustics \citep[see Fig. \ref{fig:B}, or Fig. 3 of ][]{Hui2002}.\footnote{The original figure from \citet{Hui2002}---and the motivation for understanding atmospheric refraction in exoplanetary systems---was reiterated recently by \citet{Deming2017}.} In 2002, HD 209458 b was the only known transiting exoplanet. Without a larger sample, \citet{Hui2002} could not extensively test their theory or predict the existence of atmospheric lensing in a particular system. They left a full investigation of stellar, planetary, and atmospheric composition parameters to future work. 

In the 15 years since the work of \citet{Hui2002}, 2733 additional transiting exoplanets have been discovered,\footnote{NASA Exoplanet Archive, accessed 2017 September 16.\label{test}} primarily by the \kepler\ mission. I populate Fig. \ref{fig:B} with KOIs that have published values of $a$, $R_{\star}$, $T_{\rm eq}$, and planetary radius (i.e., $z_{\rm ref}$).\textsuperscript{\ref{test}} Planetary mass estimates are found using the empirical, deterministic relations of \citet{Chen2017}, and the equilibrium temperature is used as the atmospheric temperature in all cases. An \h\ atmosphere is considered for each planet as opposed to 100\% \water, 100\% \carbon, etc. Although this assumption is not entirely realistic, it ensures that the result is an upper limit, since refraction effects are significantly smaller for high mean molecular mass atmospheres (\S\ref{sec:param_space}). 

I find the values of $\xi_{\tau=1}$, $b_{\tau=1}$, and $B_{\rm HS02}$ at visible wavelengths for each of the 82944 exoplanet systems in the parameter space, where Rayleigh scattering is the source of opacity. Fig. \ref{fig:contour_full} demonstrates that the variation in the strength of refraction across parameter space is smooth, so the five-dimensional grid is linearly interpolated to calculate $\xi_{\tau=1}$, $b_{\tau=1}$, and $B_{\rm HS02}$ for each KOI. The results are shown in Fig. \ref{fig:B}. Any parameters with values outside of the grid are assigned to the nearest node. This mostly applies to extremely short-period exoplanets, where the effects are refraction are unimportant regardless. 

According to Eq. \ref{eq:strong_lens}---the strong lensing condition of \citet{Hui2002}---a significant fraction of KOIs (residing above the black line in Fig. \ref{fig:B}) display caustics, suggesting that refraction effects dominate over Rayleigh scattering. Since the vast majority of KOIs have $a/R_{\star}<100$, this disagrees with the conditions required for significant out-of-transit refracted light signals described in \S\ref{sec:param_space} and \S\ref{sec:lens}. The discrepancy potentially arises from the mathematical formalism employed by \citet{Hui2002}, which is founded on the canonical lens equation from gravitational lensing theory \citep[e.g.,][]{Paczynski1986}. For gravitational lensing, the source---which is the host star in the exoplanet application---is treated as a point source and the magnification is not related to the finite value of $R_{\star}$. This fact is also evident in the condition for strong lensing. Combining Eqs. \ref{eq:B} and \ref{eq:strong_lens} yields $\xi_{\tau=1}(a/b_{\tau=1})>1$. The interpretation of this expression is a balance between the refraction angle required to deflect light around the planet ($a/b_{\tau=1}$) and that which can be achieved in the optically thin portion of the planetary atmosphere ($\xi_{\tau=1}$). However, in transiting (or non-transiting) exoplanet applications, most rays have to cross the disk of the star before being refracted by the far limb of the planet. The required bending angle in the exoplanet case is therefore $\sim a/R_{\star}$. When Eq. \ref{eq:strong_lens} is multiplied by a factor equalling the median value of $R_{\star}/b_{\tau=1}$ for all KOIs, the new caustic condition (Fig. \ref{fig:B}, red line) is more consistent with the detectability of refracted light presented throughout this work.

It is worth noting that \citet{Hui2002} do consider the host star as a collection of point sources in order to calculate transit light curves. However, this consideration is not included in their derivation of Eq. \ref{eq:strong_lens}.

Based on \citet{Hui2002} and Eq. \ref{eq:strong_lens}, I propose the following condition for the importance of out-of-transit refracted light signals in exoplanet observations:

\begin{equation}\label{eq:ref_cond}
\left (\frac{a}{R_{\star}} \right) \xi_{\tau=1} > 1  \,.
\end{equation}

\begin{figure}
\centering
\includegraphics[width=\columnwidth]{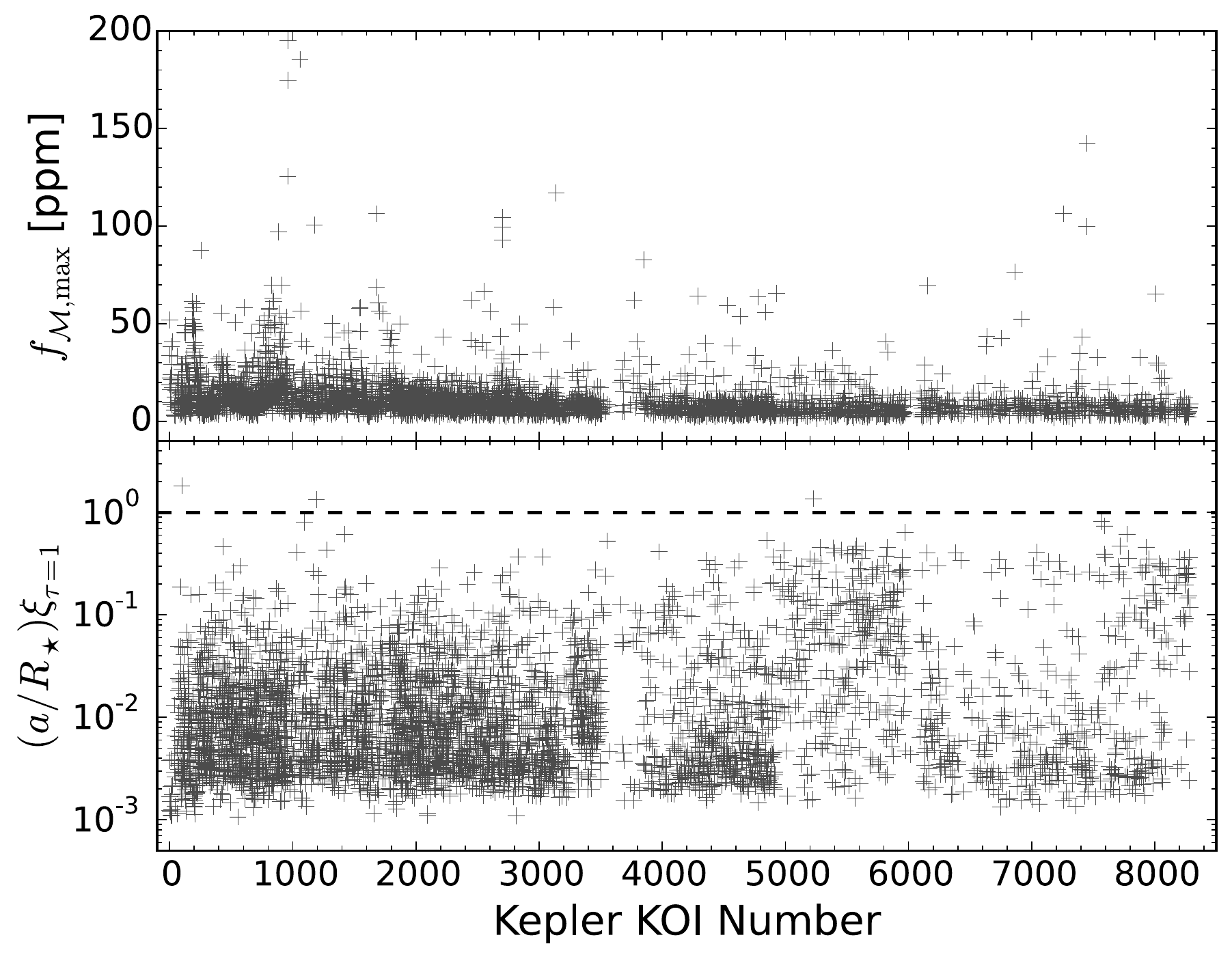}
\caption{Assessment of the detectability of atmospheric lensing by KOIs. {\bf Top:} The maximum relative flux increase ($f_{\rm \mathcal{M},max}$) as modeled in \retro\ assuming clear \h\ atmospheres. The largest signals come from short-period, hot exoplanets. {\bf Bottom:} Very few KOIs satisfy the condition for atmospheric refraction to be significant in exoplanet observations (Eq. \ref{eq:ref_cond}, dashed line). Both $f_{\rm \mathcal{M},max}$ and $(a/R_{\star}) \xi_{\tau=1}$ are needed to quantify the significance of refracted light signals in exoplanet observations. It is unlikely that individual refracted light signals are clearly distinguishable in the \kepler\ data set.}
\label{fig:KOI}
\end{figure}

\noindent As discussed above, this is an informative balance between the required refraction angle defined by the host star and orbital distance, and that which can be achieved in the optically thin portion of the exoplanet atmosphere. Fig. \ref{fig:KOI} shows that only four KOIs satisfy Eq. \ref{eq:ref_cond}: 1192.01, 5528.01, 99.01, and 1174.01 (which is out of the frame). None are confirmed exoplanets. KOIs 99.01 and 1174.01 have been identified previously as candidate long-period exoplanet systems \citep{Uehara2016,ForemanMackey2016}. KOI 5528.01 is potentially a super-Earth sized exoplanet on a $\sim$200-day orbit, and KOI 1192.01 is likely a false positive detection of an eclipsing binary.

For perspective, I also show the maximum relative flux increase $f_{\rm \mathcal{M},max}$ for all KOIs in Fig. \ref{fig:KOI}. These values are found for all KOIs by linearly interpolating the grid in the same fashion as described previously. Most of the large $f_{\rm \mathcal{M},max}$ values occur for hot exoplanets on short orbits, emphasizing that the maximum relative flux increase alone is not sufficient to quantify the significance (or lack thereof) of refracted light signals in exoplanet observations. Both $(a/R_{\star}) \xi_{\tau=1}$ and $f_{\rm \mathcal{M},max}$ are needed to understand if an observation of refracted light is feasible.

\subsection{Refraction at Long Wavelengths}\label{sec:wavelength}

The dearth of KOIs that satisfy the refraction condition (Eq. \ref{eq:ref_cond}) implies that a distinguishable individual refracted light signal is unlikely to exist in the light curve of any known or suspected exoplanet in the \kepler\ sample. This is largely a result of the short-period bias of the confirmed \kepler\ exoplanets. 

The prospect of a \emph{non-transiting} atmospheric lensing event existing in the \kepler\ data set is also rather bleak (\S\ref{sec:lens}). This is primarily a result of the substantial Rayleigh scattering opacity in the visible wavelengths spanning {\it Kepler's} bandpass. Unlike the short-period exoplanet bias, this problem has a simple solution: observe at longer wavelengths. 

All of the modeling in this work is conducted at the visible wavelength of 589.3 nm. At longer wavelengths, the Rayleigh scattering cross section decreases as $\lambda^{-4}$. Less opacity potentially enables observations of stellar mirages for many exoplanets with $a<6$ AU. Indices of refraction also decrease with increasing wavelength (e.g., Fig. \ref{fig:ior}). Since $n-1 =\nu \propto \xi$, bending angles decrease at longer wavelengths, but roughly as $\lambda^{-2}$ according to Cauchy's formula (\S\ref{sec:param_space}). The simplest estimate of the detectability of stellar mirages therefore scales as approximately $\lambda^2$. 

The combination of the these two wavelength dependencies suggests a greater likelihood of detecting a stellar mirage (or other refracted light signal) at redder wavelengths than visible. The {\it Transiting Exoplanet Survey Satellite} ({\it TESS}), poised for launch in 2018, has a notably redder bandpass than \kepler\ \citep{Ricker2015}. The ratio of their central wavelengths is roughly (800 nm)/(600 nm) $\approx$ 1.33, corresponding to an increase in detectability of $\sim$1.8. A careful estimate of the detectability of refracted light signals by {\it TESS} is beyond the scope of this work. However, the increased detectability combined with the long temporal baselines of {\it TESS'} continuous viewing zone may set the stage for a detection.

In the near-infrared portion of the spectrum, Rayleigh scattering typically yields to other sources of opacity including CIA, \water, and CH$_4$. Based on the solar system planets, the atmospheres of cold long-period gas giants are likely dominated by CH$_4$ chemistry. Fortunately, CH$_4$ has multiple 1--5 $\mu$m ``windows'' in opacity, where the absorption cross section decreases sharply. Refraction effects dominate within these windows \citep{Dalba2015}, introducing the potential for observations of refracted light with the {\it James Webb Space Telescope} ({\it JWST}).

\subsection{Science in Refracted Light}

Much of the previous literature pertaining to exoplanetary applications of refraction aims to estimate and correct for its influence on transmission spectroscopy. There have yet to be substantial efforts devoted to understanding how refraction related phenomena can be exploited to learn about exoplanets and their atmospheres \citep{Deming2017}. 

In the simplest case, a photometric detection of a stellar mirage or out-of-transit refracted light signal is illuminating. It is a method of detecting long-period exoplanets that can undergo subsequent atmospheric characterization.\footnote{Out-of-transit \emph{scattered} light is also a means of planet detection and characterization \citep{DeVore2016}. However, the scattered light signal is strongest for short-period exoplanets \citep{Robinson2017}.} Many long-period exoplanets have been discovered by radial-velocity and gravitational microlensing observations, but these techniques do not provide a method of subsequent atmospheric characterization. Directly imaged exoplanets can be characterized, but this technique is not yet sensitive enough for mature cold planets akin to Jupiter or Saturn. The sample of known long-period transiting exoplanets is small because the transit probability is low and their transits are infrequent. Even if a long-period exoplanet is detected and confirmed, high-risk observing campaigns are usually required to refine its transit ephemeris before follow-up characterization \citep[e.g., Kepler-421b][]{Kipping2014a,Dalba2016}. Refracted light offers a means of exoplanet detection that can at least partially offset some of the difficulties associated with discovering long-period exoplanets that can be characterized. In the coming decade, PLATO---an M-class mission from the European Space Agency---will locate long-period exoplanets orbiting Sun-like stars \citep{Ragazzoni2016}. PLATO is currently designed to observe at visible wavelengths, so a direct detection of refracted light in an exoplanetary system will be challenging. However, the long-period sample of exoplanets discovered by PLATO may indeed be amenable to atmospheric characterization with refracted light at redder wavelengths.

Even more information can be gleaned from a high-precision photometric observation of an atmospheric lensing event (e.g., Fig. \ref{fig:lens-crescent_rot}). From Eq. \ref{eq:ref_cond} and ray tracing models, the maximum relative flux increase informs the average refraction angle, and therefore the bulk atmospheric refractivity. This fundamental quantity is generally set by the most abundant species, which are H$_2$ and He for cold gas giant planets. Refractivity measurements are thereby complimentary to the trace abundances derived from atmospheric retrievals of transmission spectra. Helium is a notoriously difficult species to measure remotely \citep[e.g.,][]{Conrath2000}. Yet, its atmospheric abundance---as potentially revealed through refractivity measurements---greatly informs the interior evolution of a gas giant planet \citep[e.g.,][]{Fortney2003}. 

In favorable refraction conditions, an out-of-transit \emph{refraction spectrum} can be acquired with instruments such as the Wide Field Camera 3 (WFC3) on the {\it Hubble Space Telescope} ({\it HST}) and the Near Infrared Camera (NIRCAM) or Near Infrared Spectrograph (NIRSpec) that will be flown on {\it JWST}. A refraction spectrum would resemble the stellar spectrum, although wavelengths corresponding to atmospheric opacity would be attenuated. The wavelengths of transmission indirectly reveal the absorbing species in the atmosphere. The amount of transmission in the clear regions constrain the refractivity profile and the abundances of species such as H$_2$. Further consideration of ``refraction spectroscopy'' is needed, but it could potentially improve measurements of atmospheric abundances when combined with traditional transmission spectroscopy.

\section{Conclusions}\label{sec:conclusions}

Before or after a transit, an exoplanetary atmosphere refracts light into the line of sight of a distant observer. The light creates a secondary image of the host star in the planetary atmosphere---a \emph{stellar mirage}. In a transit light curve, the unresolved stellar mirage produces a small increase in flux peaked at the moment before or after transit.

I developed a ray tracing model (\retro) to simulate out-of-transit refracted light in visible light transit observations. This model offered improvements in accuracy over previous works that considered the same effect of refraction, specifically pertaining to the shape of the stellar mirage and the corresponding increase in flux. With \retro, I conducted an investigation of the strength of refracted light signals across a comprehensive exoplanet parameter space (Fig. \ref{fig:contour_full}). The primary findings are as follows:

\begin{enumerate}
    \item In agreement with previous work, atmospheric refraction is insignificant for any type of exoplanet on a close-in orbit ($a \lesssim$ 0.2 AU). For visible light observations, Rayleigh scattering attenuates light before it reaches densities large enough to generate substantial ray bending. Between $\sim$0.2 and $\sim$6 AU, small pockets of parameter space (typically occupied by large planets orbiting small stars) are amenable to producing stellar mirages that are not hidden by Rayleigh scattering. By 10 AU, most gas giant exoplanets transiting stars similar to or smaller than the Sun display clear stellar mirages. In general, out-of-transit refraction signals are important if $(a/R_{\star})\xi_{\tau=1}>1$ is satisfied (Eq. \ref{eq:ref_cond}). This relation is a balance between the required refraction angle defined by the host star and the semi-major axis and that which can be achieved in the optically thin portion of the atmosphere. Of the entire sample of \kepler\ Objects of Interest, only three potential exoplanets (KOI 99.01, 1174.01, and 5528.01) satisfy this condition. 
    \item The relative flux increases caused by atmospheric refraction vary from zero to several hundreds of parts-per-million  depending on stellar, orbital, planetary, and atmospheric properties. Realistic \h\ atmospheres produce signals on the order of tens to a hundred parts-per-million, which are potentially observable. The corresponding signals for high mean molecular weight atmospheres are significantly lower.
    \item Out-of-transit atmospheric lensing events (Fig. \ref{fig:lens-crescent_rot}) for non-transiting exoplanets are generally more observable than the equivalent phenomenon for transiting exoplanets. I describe the fundamental signal-to-noise ratios that are associated with atmospheric lensing events and investigate the potential for such a signal to exist in the \kepler\ data set. Due to Rayleigh scattering, only small, relatively impractical portions of exoplanet parameter space yield clear lensing events at visible wavelenths (Fig. \ref{fig:detect}), and these events have integrated signal-to-noise ratios between 3 and 10. 
    \item Out-of-transit refraction phenomena including stellar mirages are best observed at longer wavelengths, since the Rayleigh scattering cross section scales as $\lambda^{-4}$ compared to refractivity which scales roughly as $\lambda^{-2}$.
\end{enumerate}
     
The potential for refracted light to elucidate the nature of exoplanetary atmospheres has not yet been realized. As future instrumentation---some of which has sensitivity in the near infrared---develops, the first detection of refraction by an exoplanetary atmosphere undoubtedly draws closer.

\acknowledgments

I acknowledge my PhD advisor Philip Muirhead for helpful conversations that guided the direction of this research. I also acknowledge Sara Seager for a useful discussion about exoplanet atmospheric refraction. I finally thank the anonymous referee for a thorough and helpful review of this work.

This research made use of the NASA Exoplanet Archive, which is operated by the California Institute of Technology, under contract with the National Aeronautics and Space Administration under the Exoplanet Exploration Program. 

This research also made use the Boston University Shared Computing Cluster located at the Massachusetts Green High Performance Computer Center.

\end{document}